\pdfoutput=1
\documentclass[12pt]{scrartcl}

\usepackage[british]{babel}
\usepackage{hyphenat}
\usepackage{setspace}
\usepackage{amsmath} 
\usepackage{mathtools}
\usepackage{amssymb}
\usepackage{mathrsfs}
\usepackage[mathscr]{eucal}
\usepackage{physics}
\usepackage{bm}
\usepackage{slashed}
\usepackage{extarrows}

\usepackage{tikz}
\usetikzlibrary{
    cd, 
    arrows, 
    shapes, 
    calc, 
    positioning, 
    decorations.pathmorphing, 
    decorations.pathreplacing, 
    shapes.misc, 
    decorations.markings,
    external
}

\tikzset{->-/.style={decoration={markings, mark=at position .5 with {\arrow{>}}}, postaction={decorate}}}
\tikzset{-<-/.style={decoration={markings, mark=at position .5 with {\arrow{<}}}, postaction={decorate}}}

\usepackage{atbegshi}

\usepackage{jheppub}
\DeclareOldFontCommand{\tt}{\normalfont\ttfamily}{\mathtt}

\author{Vladimir Bashmakov}

\affiliation{Institute for Molecules and Materials, Radboud University, Heijendaalseweg 135,6525AJ\\ Nijmegen, The Netherlands}

\emailAdd{v.bashmackov@gmail.com}   

\abstract{In this note we explore monodromy defects for non-invertible symmetries in Maxwell theory, exploiting the conformal mapping to $AdS_3\times S^1$. With this approach we recover the spectrum of the defect conformal primaries. We also dedicate some time discussing the behaviour of Wilson/'t Hooft lines in the presence of such a monodromy defect, and highlight the following aspects of their behaviour: i) the lines can terminate on the defect, ii) lines of the unit electric (magnetic) charge may seize to be indecomposable, and can be represented as integer powers of some more elementary lines, and iii) they behave as topological objects when brought close to the defect, and this behaviour is governed by a Chern-Simons theory.}

	\makeatletter
	\newcommand{\aextp}{\@ifnextchar^\@aextp{\@aextp^{\,}}}
	\def\@aextp^#1{\mathop{\bigwedge\nolimits^{\!#1}}}
	\makeatother
	
	\makeatletter
	\newcommand{\extp}{\@ifnextchar_\@extp{\@extp_{\,}}}
	\def\@extp_#1{\mathop{\aextp\nolimits_{\!#1}}}
	\makeatother

\preprint{}

\title{Monodromy Defects for Electric-Magnetic Duality, Hyperbolic Space, and Lines} 

\newcommand{\beq}{\begin{equation}}
\newcommand{\eeq}{\end{equation}}
\newcommand{\bea}{\begin{eqnarray}}
\newcommand{\eea}{\end{eqnarray}}

\begin{document}

\maketitle

\section{Introduction}

Conformal defects are extended operators in conformal field theory (CFT), preserving a subgroup of a conformal group that contains the scale transformation \cite{Billo:2013jda, Billo:2016cpy}. More precisely, a planar $p$-dimensional defect in a $d$-dimensional CFT preserves the $SO(p+1,1)\times SO(d-p)\subset SO(d+1, 1)$ subgroup.

Naturally, the conformal defects in free field theory allow for a more detailed treatment. In \cite{Lauria:2020emq, Nishioka:2021uef, Bianchi:2021snj, Bashmakov:2024suh, Kim:2025tvu, Bartlett-Tisdall:2025iqx} defects in free scalar field theory, with various bulk and defect dimensions, were discussed. In particular, in \cite{Lauria:2020emq} certain no-go theorems for such defects were formulated, such as the triviality of non-monodromy surface defects in $4d$ free scalar field theory. In \cite{Sato:2021eqo, Bianchi:2021snj} defects in free fermion theories were investigated.

Another natural example to look at is Maxwell theory. Line defects are well-known and are represented by the Wilson-'t Hooft lines \cite{Kapustin:2005py}. While non-monodromy surface defects in Maxwell theory have recently been proven to be trivial \cite{Herzog:2022jqv}, monodromy defects remain a viable possibility.

A monodromy defect is a codimension-two defect such that a symmetry transformation is performed while going around it: they are therefore tightly related to the global symmetries of a theory. The simplest option in the context of Maxwell theory would be the charge conjugation symmetry $\mathcal{C}$\footnote{The defects for charge conjugation have recently been considered in \cite{Gomis:2025gzb} in the context of $\mathcal{N}=4$ SYM.}: the charge conjugation symmetry, and hence the corresponding defect, exists for any value of the complexified coupling constant $\tau = \frac{2\pi\,i}{e^2}+\frac{\theta}{2\pi}$.

Another intriguing possibility comes from the renown electric-magnetic duality enjoyed by the theory. For a generic value of $\tau$ the transformation changes the coupling,
\begin{equation}
\mathcal{S}:\quad \tau\rightarrow \,-1/\tau,
\end{equation}
and so is indeed a duality between two theories at different couplings. However, at the special value $\tau=i$ the theory at hand is left invariant, so the duality turns into a symmetry. The modern point of view on the symmetries of quantum systems suggests that symmetries are associated with topological operators \cite{Gaiotto:2014kfa}. Correspondingly, the electric-magnetic duality transformation can be implemented as a topological interface of the form \cite{Lozano:1995aq, Gaiotto:2008ak, Kapustin:2009av}
\begin{equation}\label{Eq:Sinterface}
    e^{\frac{i}{2\pi}\int_\Sigma\,A d \tilde{A}}.
\end{equation}
This is a Chern-Simons-like interaction that couples the guage fields on both sides of the interface, and $\Sigma$ is a three-manifold supporting the interface.

As was shown in \cite{Choi:2021kmx, Choi:2022zal}, we can construct a topological interface for any $\tau = i N$, $N\in\mathbb{N}$. While the electric-magnetic duality transformation $\mathcal{S}$ changes the coupling, we can compensate the change by half-space gauging of the $\mathbb{Z}_{N}^{(1)}$ subgroup of the  $U(1)_e^{(1)}$ electric one-form symmetry:
\begin{equation}
    \tau=i\,N\xrightarrow{\quad\mathcal{S}\quad}\frac{i}{N}\xrightarrow{\mathbb{Z}_N-\text{gauging}} i\,N.
\end{equation}
The resulting topological interface is a generalization of \eqref{Eq:Sinterface},
\begin{equation}\label{Eq:SNinterface}
    \mathcal{D}_N^{(2)}=e^{\frac{iN}{2\pi}\int_\Sigma\,A d \tilde{A}},
\end{equation}
and the resulting symmetry happens to be \textit{non-invertible}, that is, fails to have an inverse \cite{Choi:2021kmx, Choi:2022zal, Schafer-Nameki:2023jdn, Shao:2023gho, Kaidi:2026urc}. Clearly, a similar situation also takes place for $\tau = i/N$, $N\in\mathbb{N}$, with the $\mathbb{Z}_{N_m}^{(1)}$ subgroup of the magnetic $U(1)_m^{(1)}$ being gauged instead.

Even more generally, one can show that there are topological defects for any rational point $\tau=\frac{i\,N_e}{N_m}$, $N_e, N_m\in\mathbb{N}$, $\text{gcd}(N_e,\,N_m)=1$ \cite{Niro:2022ctq, Cordova:2023ent}. To construct such defect, one has to concatenate the $\mathcal{S}$ transformation with the half-space gauging of $\mathbb{Z}_{N_e}^{(1)}\times\mathbb{Z}_{N_m}^{(1)}\in U(1)_e^{(1)}\times U(1)_m^{(1)}$, and the corresponding interface is given by
\begin{equation}\label{Eq:DualityRationalPoint}
    \mathcal{D}_{N_e, N_m}^{(2)}=e^{\frac{i}{2\pi}\int_{\Sigma}\,N_m\,a\,db-N_e a\,dA+b\,d\tilde{A}}.
\end{equation}

Besides the $\mathcal{S}$-duality transformation, one can also consider other elements of the $SL(2,\mathbb{Z})$ duality group, such us the $\mathcal{ST}$ transformation, where $\mathcal{T}:\,\tau\rightarrow\tau+1$. This is a third order operation (as seen from the action on the coupling constant)\footnote{ And a sixth order operation when acting on the fields, $(\mathcal{S}\mathcal{T})^3=\mathcal{C}$.}, and one can construct triality defects. More precisely, for $\tau = e^{2\pi i/3}N$ ($N\in\mathbb{N}$) one can have a defect with the action
\begin{equation}\label{Eq:Duality_Fractional}
    \mathcal{D}_{N}^{(3)}=e^{\frac{i}{4\pi}\int_{\Sigma}\,2N\,A\,d\tilde{A}-NA\,dA}.
\end{equation}

In fact, the general analysis of \cite{Niro:2022ctq} revealed that similar, generically non-invertible, defects involving electric-magnetic duality transformations and half-space gauging of discrete subgroups of the one-symmetry group can be constructed for any rational value of $\tau$.

The goal of this note is to consider monodromy defects for these non-invertible symmetries, which one can also think of as boundaries of the corresponding topological defects. In doing that, we are going to conformally map the system from $\mathbb{R}^4$ to $AdS_3\times S^1$; then many properties of the defect can be read off, using the standard holographic dictionary.

Upon the $S^1$ reduction, one obtains an effective theory in $AdS_3$, which consists of a tower of massive Kaluza-Klein modes and also contains a topological sector; one can say that this effective $3d$ theory is non-trivially gapped. The topological sector is responsible for an extra sector of the defect conformal primaries. Moreover, it governs the large-distance behaviour of the line operators, which is translated to the near-the defect limit while coming back to $\mathbb{R}^4$.

In the remainder of this section we give a more precise formulation of the setup, using the $\mathcal{D}_N^{(2)}$ defect as an example.

\subsection{Monodromy for duality defects}

We consider Maxwell theory in flat Minkowski space-time with the action
\begin{equation}
    S_{\text{Max}} = -\int \frac{1}{2e^2}\mathcal{F}\wedge\star\mathcal{F}+\frac{\theta}{8\pi^2}\mathcal{F}\wedge\mathcal{F},
\end{equation}
and introduce a cut along the $\Sigma = (x>0, y=0)$ semi-plane. We will then supplement the theory with the surface term, defined on this semi-plane:
\begin{equation}
    S_{\text{cut}} = \frac{N}{2\pi}\int_{\Sigma}\,A_L\wedge dA_U.
\end{equation}
Here, $A_U$ and $A_L$ are the values of the gauge field on the upper edge of the cut and on the lower edge of the cut, respectively.
\begin{figure}
\begin{center}
    \begin{tikzpicture}[scale=1,>=stealth]
    
    \draw[line width =1pt, color=green!55!blue] (0,0) -- (4, 0);
    \filldraw[blue!55!green] (0, 0) circle (2pt);
    \node[right] at (-0.25, -0.25) {\text{ $\mathcal{M}$}};
    \node[above] at (2, 0) {\text{ $A_U$}};
    \node[below] at (2, 0) {\text{ $A_L$}};

    \draw[line width = 1pt,  color = black, postaction={decorate}, decoration={markings, mark=at position 0.5 with {\arrow[scale=1.5]{>}}}] (0, 0) circle (20pt);

    \node[left] at (-0.8, 0) {\text{ $\mathcal{D}_N^{(2)}$}};

    \node[below] at (3.8, 0) {\text{ $\Sigma$}};
    
    \end{tikzpicture}

    \caption{Monodromy defect $\mathcal{M}$, a boundary of the topological defect $\mathcal{D}^{(2)}_N$ supported on the surface $\Sigma$.}
    \label{fig:Monodromy}

\end{center}
\end{figure}
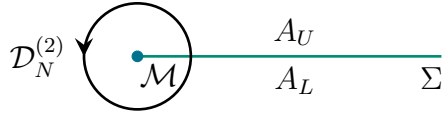
We also have a freedom to add a purely defect counterterm; the standard boundary counterterm for the Abelian Chern-Simons theory, accomplishing a well-defined variational problem, is
\begin{equation}
    S_{\text{CT}}=-\frac{N}{2\pi}\int_{\partial\Sigma} d^2zA_zA_{\bar{z}},
\end{equation}
which is defined provided that the choice of a complex structure on $\partial\Sigma$ was made.

The equation of motion for the gauge fields derived by varying the total action (here we put $\theta=0$) $S_{\text{tot}}=S_{\text{Max}}+S_{\text{cut}}+S_{\text{bdry}}$. The variation is given by
\begin{eqnarray}
    \delta S_{\text{EM}} &=& -\int\,\frac{1}{e^2}\delta A\wedge d\star\,\mathcal{F}\,+\,\int_{\Sigma} \left(-\frac{1}{e^2}\delta A_{U}\wedge\star\mathcal{F}_U + \frac{1}{e^2}\delta A_{L}\wedge\star\mathcal{F}_L \right),\nonumber\\
    \delta S_{\text{cut}} &=& \frac{N}{2\pi}\int_{\Sigma}\,\left(\delta A_{U}\wedge\mathcal{F}_L+\delta A_{L}\wedge \mathcal{F}_U\right)+\frac{N}{2\pi}\int_{\partial\Sigma}A\wedge\delta A,\nonumber\\
    \delta S_{\text{CT}}&=&-\frac{N}{2\pi}\int_{\partial\Sigma}\,\delta A_zA_{\bar{z}}+A_z\delta A_{\bar{z}}.
\end{eqnarray}
Requiring the vanishing of the boundary terms on $\Sigma$, and putting the coupling $\tau \equiv \frac{2\pi i}{e^2}=Ni$, we find the boundary conditions
\begin{equation}\label{eq:DualityGluing}
    \mathcal{D}_N^{(2)}:\quad\left(\mathcal{F}_U + \star\mathcal{F}_L\right)\vert_{\Sigma}=0,\quad \left(\mathcal{F}_L -\star\mathcal{F}_U\right)\vert_{\Sigma}=0,
\end{equation}
which can also be rewritten as
\begin{equation}\label{S-monodromy}
    \mathcal{D}_N^{(2)}:\quad\mathcal{F}(\phi=2\pi) = \star\mathcal{F}(\phi=0).
\end{equation}

Considering the $\mathcal{D}_{N_e, N_m}^{(2)}$ defect with $\tau=\frac{2\pi i}{e^2}=\frac{N_e}{N_m}i$, one finds the same gluing condition \eqref{eq:DualityGluing}. On the other hand, the gluing condition for the triality defect $\mathcal{D}_N^{(3)}$ takes the form
\begin{equation}
    \mathcal{D}_N^{(3)}:\qquad \mathcal{F}(\phi=2\pi) = \frac{\sqrt{3}}{2}\star\mathcal{F}(\phi=0)+\frac{1}{2}\mathcal{F}(\phi=0).
\end{equation}

\paragraph{On the $1$-form symmetry breaking} The variational term supported on $\partial\Sigma$ leads to a modification of the Maxwell equation:
\begin{equation}
    \left(\frac{d\star\mathcal{F}}{e^2}\right)_{\rho\phi\bar{z}}=\frac{N}{\pi}A_{\bar{z}}\wedge\delta(\partial\Sigma),\qquad
    \left(\frac{d\star\mathcal{F}}{e^2}\right)_{\rho\phi z}=0.
\end{equation}
$\frac{1}{e^2}\star\mathcal{F}$ can be recognized as the conserved current for the $U(1)^{(1)}_e$ electric one-forms symmetry, and the above equation implies that this symmetry is broken at the defect position. Similar breaking also occurs for the magnetic $U(1)^{(1)}_m$.

In general, consider a conformal defect breaking a one-form symmetry conserved current $J_{\mu\nu}$:
\begin{equation}\label{Eq:GeneralOneFormBreaking}
\partial^{\mu} J_{\mu\nu} = j_{\nu}\,\delta^{\perp}(0).   
\end{equation}
It is natural to separate the tangential component $j_{i}$, which is a vector operator on the defect, and the normal component $j_a$, which is a scalar on the defect charged under the rotations around the defect. The tangential component has the dimension $p-1$, where $p$ is the defect dimension, and thus by unitarity should be a conserved current. Applying $\partial^{\nu}$ to \eqref{Eq:GeneralOneFormBreaking}, using the antisymmetry of $J_{\mu\nu}$ and the conservation of $j_{i}$, we find:
\begin{equation}
    0=j_a\,\partial^a \delta^{\perp}(0),
\end{equation}
implying that $j_a$ vanishes and there is only the tangential part on the RHS of \eqref{Eq:GeneralOneFormBreaking}. It can be anticipated that $A_{\bar{z}}$ will play the role of such a (chiral) current living on the monodromy defect.

We can give a simple interpretation to this result. The defect, breaking a $U(1)^{(1)}$ one-form symmetry, will enjoy a zero-form $U(1)^{(0)}$ symmetry with some defect operators being charged under it. These charged operators can serve as the end points of the bulk lines, charged under the broken $U(1)^{(1)}$.

\section{AdS Mapping}
We have already stated that monodromy defects are natural candidates for conformal defects. More precisely, one may consider a class of defects implementing a given monodromy transformation, and conclude that they should flow to a non-trivial conformal defect(s) in the IR.\footnote{The symmetry transformation can be observed at arbitrary long distance, and therefore the defect can not flow to a trivial one.} A useful point of view on conformal defects comes from the conformal mapping to $\text{AdS}_{p+1}\times S^{d-p-1}$ \cite{Kapustin:2005py, Gadde:2016fbj, Hung:2014npa, Nishioka:2021uef, Giombi:2021uae}. In this picture the defect operators manifest themselves as the boundary operators. The bulk fields of the original flat space theory reduced on $S^{d-p-1}$ give rise to Kaluza-Klein towers in $\text{AdS}_{p+1}$, and these massive fields are related to the boundary operators via the usual holographic dictionary.  

For the case at hand, we conformally map the problem to $AdS_3\times S^1$, with the metric
\begin{equation}
    ds^2=\ell^2\frac{-dr^2+d\vec{y}^2}{r^2}-\ell^2d\theta^2,
\end{equation}
where $\ell$ is the radius of $AdS_3$ and $S^1$, and $S^1$ is twisted by the $\mathcal{D}$ transformation. Following the standard procedure, we reduce the model on the circle and obtain a tower of Kaluza-Klein modes in $3d$. In fact, reducing a gauge theory down to three dimensions, it is natural to expect the appearance of massive modes along with a topological sector, associated with flat connections. It is convenient to discuss these two sectors separately, and we start from the former.

\subsection{Massive sector.} Let us start with the $\mathcal{D}^{(2)}$ case. The Maxwell equations in four dimensions read
\begin{equation}\label{Gluing condition}
    d\mathcal{F}=0,\qquad d\star\mathcal{F}=0.
\end{equation}
At some point on the circle, assumed to be $\theta=0\sim\theta=2\pi$, we have a cut, and the boundary conditions along the cut are
\begin{eqnarray}
    \mathcal{F}(2\pi) = \star\mathcal{F}(0).
\end{eqnarray}
It is convenient to decompose the $4d$ curvature two-form $\mathcal{F}$ as $\mathcal{F}=F+B\wedge d\theta$, where $F$ is a $3d$ curvature associated with a three-dimensional Abelian gauge field, and $B$ is a $3d$ vector field which can be thought of as the "curvature" associated with a compact boson. The Maxwell equations take the form
\begin{eqnarray}
    &&dF+\partial_{\theta}F\wedge d\theta+dB\wedge d\theta=0,\\
    &&-\ell^{-1}d\star B-\ell^{-1}\star\,\partial_{\theta}B\wedge d\theta + \ell\,d\star F\wedge d\theta=0,
\end{eqnarray}
where both the exterior derivative and the star operation are understood as three-dimensional ones. These equations take a more symmetric form ones the vector $B$ is dualized via $G=\ell^{-1}\star\,B$, which corresponds to dualizing a compact scalar to an Abelian gauge field. This yields the equations
\begin{eqnarray}\label{Eq:3DEOMs}
    &&d\star F-\ell^{-1}\partial_{\theta}G=0,\\
    &&d\star G+\ell^{-1}\partial_{\theta}F=0,\\
    &&dF=0,\\
    &&dG=0.
\end{eqnarray}
The glueing condition \eqref{Gluing condition} takes the form
\begin{equation}\label{Eq:DualityB.c.}
    F(2\pi)=-G(0),\qquad G(2\pi)=F(0).
\end{equation}
Note that while we impose this relation at a single point $\theta=0\sim\theta=2\pi$, due to the equations of motion, it holds at any point: $F(\theta+2\pi)=-G(\theta)$, $G(\theta+2\pi)=F(\theta)$.

We now have to perform the twisted reduction on $S^1$. It is useful to define the object $H=F+iG$ and its complex conjugate, which are just multiplied by $\pm i$ under the action of electric-magnetic duality, $H \xlongrightarrow{\mathcal{S}} i H$. We then get the equation
\begin{equation}
    d\star\,H+i\ell^{-1}\partial_{\theta}H=0,
\end{equation}
and expand $H$ into angular modes,
\begin{equation}
    H\,=\,\sum_{n=-\infty}^{\infty}\,H_n\,e^{i\theta(n+1/4)}.
\end{equation}
The modes $H_n$ depend on the coordinates on $\text{AdS}_3$, and in turn satisfy the equations
\begin{equation}
    d\star\,h_n-\ell^{-1}\left(n+\frac{1}{4}\right)h_n=0.
\end{equation}
This is equivalent to two Maxwell-Chern-Simons (MSC) equations, for real and imaginary parts of $h_n$, respectively. The mode has positive (negative) chirality, if $n+\frac{1}{4}>0$ ($n+\frac{1}{4}<0$). Note that there are no massless propagating gauge fields in the bulk: the $\mathcal{S}$ duality twist gaps the spectrum in $AdS_3$. On the other hand, we find a massive gauge field for each $n\in\mathbb{Z}$, with masses $m_n=\ell\, |n+1/4|$ and $m_{-n}=\ell\,|n+3/4|$, where now $n\in\mathbb{Z}_{\geq0}$. The holographic dictionary tells us that the dual operators are vector operators with the following dimensions and $SO(2)$ charges \cite{Andrade:2011sx}:
\begin{eqnarray}
    (h, \bar{h})\,&=&\,\left(1+\frac{n+1/4}{2}, \frac{n+1/4}{2}\right), \qquad q=n+1/4,\qquad n\in\mathbb{Z}_{\geq0},\\
    (h, \bar{h})\,&=&\,\left(\frac{n+3/4}{2}, \frac{n+3/4}{2}+1\right), \qquad q=n+3/4,\qquad n\in\mathbb{Z}_{\geq0}.
\end{eqnarray}
We note that these results do not depend on $N_e$ (or $N_m$), and are determined by the fact that $\mathcal{D}^{(2)}_N$ (or $\mathcal{D}^{(2)}_{N_e, N_m}$) comes from a duality transformation. This is also consistent with the findings in \cite{Shao:2025qvf}, where these generalized free field operators were referred to as the universal sector.

Along the same line, one may consider the twist with a triality transformation $\mathcal{D}^{(3)}_N$. The monodromy condition similar to \eqref{Eq:DualityB.c.} reads
\begin{eqnarray}\label{Eq:TrialityGluingb.c.}
    F(\phi=2\pi) &=& \frac{1}{2}F(\phi=0)-\frac{\sqrt{3}}{2}G(\phi=0),\nonumber\\
    G(\phi=2\pi) &=& \frac{\sqrt{3}}{2}F(\phi=0)+\frac{1}{2}G(\phi=0),
\end{eqnarray}
and the mode equation takes the form
\begin{equation}
    d\star\,h_n-\ell^{-1}\left(n+\frac{1}{6}\right)h_n=0,
\end{equation}
and the generalized free sector dimensions and charges are given by

\begin{eqnarray}
    (h, \bar{h})\,&=&\,\left(1+\frac{n+1/6}{2}, \frac{n+1/6}{2}\right), \qquad q=n+1/6,\qquad n\in\mathbb{Z}_{\geq0},\\
    (h, \bar{h})\,&=&\,\left(\frac{n+5/6}{2}, \frac{n+5/6}{2}+1\right), \qquad q=n+5/6,\qquad n\in\mathbb{Z}_{\geq0}.
\end{eqnarray}

\subsection{Line operators}

Next, we consider line operator insertions; in order to simplify the analysis, we will assume that the line is located at a point $\theta=\theta_l$ on a circle. It is natural to conclude from the construction that two types of line insertion can be distinguished: the lines that live out of the topological interface ($0<\theta_l<2\pi$), and the lines that live on the interface ($\theta_l=0$ or $\theta_l=2\pi$). We will discuss these two types of line separately, starting with the former. HOwever, it is worth noticing that by suitably deforming the shape of the cut, any bulk line can be realized as a defect line; as we will se, the opposite is not always true, and certain lines are intrinsically bound to the topological defect.

A Wilson line $\mathcal{W}_n(\gamma)$ placed out of the defect appears as a source in the equations of motion,
\begin{equation}
    -\partial_{\theta}G\wedge d\theta + \ell\,d*F\wedge d\theta=e^2\,\delta(\gamma)\wedge\delta(\theta-\theta_l).
\end{equation}
The localized source can also be recast in terms of gluing conditions at $\theta_l$,
\begin{equation}\label{Eq:BulkWilsonb.c.}
    F(\theta_l+\epsilon)=F(\theta_l-\epsilon),\qquad G(\theta_l+\epsilon)=G(\theta_l-\epsilon)-e^2\delta(\gamma).
\end{equation}

Next, we consider an 't Hooft line; again, for simplicity we will consider the line located at some fixed angle position $\theta_l$. By definition, insertion of an 't Hooft line along the contour $\gamma$ prescribes the flux alone any homology two-sphere, linked with $\gamma$ with the linking number one:
\begin{equation}
    \int_{S^2}\,\mathcal{F} = 2\pi \,m,\quad m\in\mathbb{Z}.
\end{equation}
Let us choose this homology two-sphere to be a cylinder $\mathcal{C}$ with the bases given by two disks located at some fixed angles $\theta_1$, $\theta_2$, $\theta_1<\theta_l<\theta_2$, parallel to each other: $\mathcal{C}= D_1\cup [\theta_1, \theta_2]\times S^1\cup D_2$. Then the flux condition gives
\begin{equation}
    \int_{\mathcal{C}}\,\mathcal{F} = \int_{D_2}F- \int_{D_1}F+\int_{[\theta_1,\theta_2]\times S^1}B\wedge d\theta = 2\pi\,m.
\end{equation}
With shrinking the size of the cylinder, the condition above translates to the gluing condition
\begin{equation}\label{Eq:BulktHooftB.c.}
 F(\theta_l-\epsilon) = F(\theta_l+\epsilon) - 2\pi\,m\,\delta(\gamma),\qquad G(\theta_l-\epsilon) = G(\theta_l+\epsilon).
\end{equation}
Notice that if we start with the gluing conditions \eqref{Eq:BulkWilsonb.c.} and go around the circle once, the $F$ and $G$ fields will swap according to \eqref{Eq:DualityB.c.}, and we will find \eqref{Eq:BulktHooftB.c.} with $m = \frac{n}{N}$. demonstrating that in the background at hands a Wilson line of charge $n$ is equivalent to an 't Hooft line of charge $\frac{n}{N}$, in agreement with \cite{Choi:2021kmx, Choi:2022zal}. We recall that, unless $N$ divides $n$, this 't Hooft line has a fractional charge, and so needs to be equipped with a surface attached for being a well-defined gauge invariant operator.

In the presence of the $\theta$-angle, due to the Witten effect, the condition \eqref{Eq:BulktHooftB.c.} is modified to
\begin{equation}
 F(\theta_l-\epsilon) = F(\theta_l+\epsilon) - 2\pi\,m\,\delta(\gamma),\qquad G(\theta_l-\epsilon) = G(\theta_l+\epsilon)-\frac{e^2\,\theta\,m}{2\pi}\delta(\gamma).
\end{equation}

We now proceed with the lines put directly on the topological defect. Their insertions modify the monodromy conditions, and the modifications are case-dependent, so we go through the examples one by one.

\paragraph{$\mathcal{D}^{(2)}_N$ defect} The insertion of the line
\begin{equation}
    \mathcal{W}_n^{\mathcal{D}}(\gamma)=e^{in\int_{\gamma}A_L}
\end{equation}
(the superscript reminds that the line lives on the interface) on the lower side of the duality defect cut modifies \eqref{Eq:DualityB.c.} to
\begin{equation}\label{Eq:DualityWilsonLowerB.c.}
    F(2\pi)=-G(0),\qquad G(2\pi)=F(0)+\frac{2\pi n_L}{N}\,\delta(\gamma_L),
\end{equation}
where $\delta(\gamma)$ is the two-form  the delta-function support Poincar\'e dual to $\gamma$. Similarly, for a line on the upper side of the cut
\begin{equation}\label{Eq:DualityWilsonUpperB.c.}
        F(2\pi)=-G(0)-\frac{2\pi n_U}{N}\,\delta(\gamma_U),\qquad G(2\pi)=F(0).
\end{equation}

We can obtain a defect Wilson line by pushing a bulk line on the defect. The result of pushing a line on the lower cut is obtained by first taking the boundary condition \eqref{Eq:BulkWilsonb.c.}, and concatenating it with the monodromy transformation \eqref{Eq:DualityB.c.}: this results in \eqref{Eq:DualityWilsonLowerB.c.}. On the other hand, by taking first the monodromy condition \eqref{Eq:DualityB.c.}, and then applying \eqref{Eq:BulkWilsonb.c.}, we recover \eqref{Eq:DualityWilsonUpperB.c.}. This can be summarized in a (pretty much natural) relations
\begin{equation}\label{Eq:WilsonBulkDefect}
    \mathcal{W}_n(\gamma)\times\mathcal{D}_N= e^{in\int_{\gamma}A_L},\qquad \mathcal{D}_N\times \mathcal{W}_n(\gamma)= e^{in\int_{\gamma}A_U}.
\end{equation}

With the help of \eqref{Eq:BulktHooftB.c.} and \eqref{Eq:DualityB.c.}, and making again comparison with \eqref{Eq:DualityWilsonLowerB.c.} and \eqref{Eq:DualityWilsonUpperB.c.}, we can push an 't Hooft line on both sides of the cut. The result is summarized in the relations
\begin{equation}\label{Eq:tHooftBulkDefect}
    \mathcal{T}_m(\gamma)\times\mathcal{D}_N= e^{-imN\int_{\gamma}A_U},\qquad \mathcal{D}_N\times \mathcal{T}_m(\gamma)= e^{imN\int_{\gamma}A_L}.
\end{equation}
In particular, by equating the RHSs of \eqref{Eq:BulktHooftB.c.} and \eqref{Eq:tHooftBulkDefect}, we can recover the familiar duality transformation rules:
\begin{equation}
    \mathcal{W}_n(\gamma)\times\mathcal{D}_N = \mathcal{D}_N\times \mathcal{T}_{\frac{n}{N}}(\gamma),\qquad\mathcal{T}_n(\gamma)\times\mathcal{D}_N=\mathcal{D}_N\times \mathcal{W}_{-nN}(\gamma).
\end{equation}

As the result, we are able to represent the Wilson/'t Hooft lines as the (Wilson) lines on the defect. In practice, having some line insertions, we can deform the symmetry defect such that the lines are now found on its surface, and hence are represented by the defect lines. This representation is not unique, since we can start deforming the defect both clockwise and counter-clockwise, fusing the lines from the left or from the right, respectively. It is also interesting to note that the 't Hooft line of minimal charge turns out to be decomposable, that is, given by some integer power of another line.

\paragraph{$D^{(2)}_{N_e,\,N_m}$ defect} The main novelty here is that while inserting lines on the duality defect, we can also consider the lines associated with the intrinsically three-dimensional fields $a$ and $b$. The most general insertion 
\begin{equation}
    n_L\int_{\gamma_L}A_L+n_U\int_{\gamma_U}A_U+n_a\int_{\gamma_a}a+n_b\int_{\gamma_b}b 
\end{equation}
leads to the gluing condition
\begin{eqnarray}
    F(2\pi)&=& -G(0)-\frac{2\pi\,n_U\,N_m }{N_e}\,\delta(\gamma_U)+\frac{2\pi\,n_a}{N_e}\,\delta(\gamma_a),\nonumber\\
    G(2\pi)&=&F(0)+\frac{2\pi\,n_L\,N_m}{N_e}\,\delta(\gamma_L)+2\pi\,n_b\, \delta(\gamma_b).
\end{eqnarray}
This enables us to write down the following fusion rules:
\begin{eqnarray}
    \label{Eq:FusionRationalPoint1}
    \mathcal{W}_n\times\mathcal{D}_{N_e,\,N_m}&:&\quad n_U\,N_m-n_a=0,\quad n_L\,N_m+n_b\,N_e=nN_m,\\
    \label{Eq:FusionRationalPoint2}
    \mathcal{D}_{N_e,\,N_m}\times \mathcal{W}_n&:&\quad n_U\,N_m-n_a=n\,N_m,\quad n_L N_m+n_b\,N_e=0,\\
    \label{Eq:FusionRationalPoint3}
    \mathcal{T}_m\times\mathcal{D}_{N_e,\,N_m}&:&\quad n_U\,N_m-n_a=-m\,N_e,\quad n_b\,N_e + n_L N_m=0,\\
    \label{Eq:FusionRationalPoint4}
     \mathcal{D}_{N_e,\,N_m}\times \mathcal{T}_m&:&\quad n_a-n_U\,N_m=0,\quad n_L\,N_m+n_b\,N_e=m\,N_e.
\end{eqnarray}
Notice that the correspondence is not one-to-one, and a given bulk line can generically be represented by a few defect lines. We can associate a $K$-matrix with the defect \eqref{Eq:DualityRationalPoint},
\begin{equation}
    K = \left(\begin{matrix}
        0 && 0 && -N_e && 0\\
        0 && 0 && 0 && 1\\
        -N_e && 0 && 0 && N_m\\
        0 && 1 && N_m && 0
    \end{matrix}\right),
\end{equation}
then the solutions differ by the image of $K$. This ambiguity can be attributed to the redefinition freedom of the fields $a$ and $b$.

The conditions \eqref{Eq:FusionRationalPoint1}- \eqref{Eq:FusionRationalPoint4} can be repackaged in the following way. Using the fact that $N_e$ and $N_m$ are coprime, we introduce two integers $p,\,q\in\mathbb{Z}$, such that $p\,N_m-q\,N_e=1$. We can then consider the lines
\begin{equation}
    l_1=e^{i\int p\,A_L - q\,b},\quad l_2=e^{i\int A_U+q\,N_e\,a}.
\end{equation}
We then have:
\begin{eqnarray}
    \mathcal{W}_n\times\mathcal{D}_{N_e,\,N_m}&=&l_1^{n\,N_m},\qquad \mathcal{D}_{N_e,\,N_m}\times \mathcal{W}_n=l_2^{n\,N_m},\label{Eq:WilsonBulkDefect2}\\
    \mathcal{T}_m\times\mathcal{D}_{N_e,\,N_m}&=&l_2^{-m\,N_e},\qquad \mathcal{D}_{N_e,\,N_m}\times \mathcal{T}_m=l_1^{m\,N_e}.\label{Eq:tHooftBulkDefect2}
\end{eqnarray}

Also in this case neither Wilson, not 't Hooft lines of minimal charge are not indecomposable lines now, but rather are given by integer powers of some other line operators. The lower cut (upper cut) Wilson line of charge one is given by $l_1^{N_m}$ ($l_2^{N_m}$). Correspondingly, the lower cut (upper cut) 't Hooft line of charge one is given by $l_2^{-N_e}$ ($l_1^{N_e}$).

Comparing the RHS's of \eqref{Eq:WilsonBulkDefect2}, we recover the duality transformations
\begin{equation}
    \mathcal{W}_{r\,N_e}\times\mathcal{D}_{N_e,\,N_m}=\mathcal{D}_{N_e,\,N_m}\times \mathcal{T}_{r\,N_m},\qquad \mathcal{T}_{r\,N_m}\times\mathcal{D}_{N_e,\,N_m}=\mathcal{D}_{N_e,\,N_m}\times \mathcal{W}_{-r\,N_e}.
\end{equation}

At the level of correlators, applying the cyclic permutations inside the trace, we are also entitled to equate the results of fusing a line from below and from above (or from the left and from the right). This implies that
\begin{equation}
    \langle l_1^{nN_m}\rangle= \langle l_2^{nN_m}\rangle,\qquad \langle l_1^{nN_e}\rangle= \langle l_2^{-nN_e}\rangle.
\end{equation}
Notice that this time both the minimal charge Wilson line and the minimal charge 't Hooft line can be decomposed into more elementary line operators.

\paragraph{$\mathcal{D}^{(3)}_N$ defect} Finally, we take $e^2 = \frac{4\pi}{\sqrt{3}N}$ and $\theta = \pi\,N$, and consider lines placed on the triality defect.

The line insertions on the triality defects,
\begin{equation}
    n\int_{\gamma_L}A_L+m\int_{\gamma_U}A_U,
\end{equation}
lead to the gluing condition
\begin{eqnarray}
    &F(2\pi)& = \frac{1}{2}F(0)-\frac{\sqrt{3}}{2}G(0)-\frac{2\pi\,n_U}{N}\delta(\gamma_U),\\
    &G(2\pi)& = \frac{\sqrt{3}}{2}F(0)+\frac{1}{2}G(0)+\frac{2\pi\,n_U}{\sqrt{3}N}\delta(\gamma_U)+\frac{4\pi\,n_L}{\sqrt{3}N}\delta(\gamma_L).
\end{eqnarray}
Combining \eqref{Eq:BulkWilsonb.c.} with \eqref{Eq:TrialityGluingb.c.}, and comparing the result with the monodromy condition above, we find:
\begin{equation}
    \mathcal{W}_n\times\mathcal{D}^{(3)}_N=e^{in\int A_L},\quad \mathcal{D}^{(3)}_N\times\mathcal{W}_n=e^{in\int A_U},
\end{equation}
as expected. Similarly, for an 't Hooft line we get:
\begin{equation}\label{Eq:TrialitytHooftFusion}
    \mathcal{T}_m\times\mathcal{D}^{(3)}_N=e^{imN\int A_U},\qquad \mathcal{D}^{(3)}_N\times\mathcal{T}_m=e^{imN\int A_U-A_L}.
\end{equation}
These results are consistent with the triality transformation, as can be seen from comparing the RHSs:
\begin{equation}
    \mathcal{W}_{nn}\times\mathcal{D}^{(3)}_N=\mathcal{D}^{(3)}_N\times (\mathcal{W}_{nN}\cdot\mathcal{T}_{-n}),\qquad \mathcal{T}_m\times\mathcal{D}^{(3)}_N=\mathcal{D}^{(3)}_N\times \mathcal{W}_{mN}.
\end{equation}
\subsection{Flat sectors.} Next, we discuss the dynamics of the flat sector and its reduction on a circle. In order to understand the TFT governing the flat sector dynamics, let us consider a boundary with some generic topology, e.g. a Riemann surface $\Sigma_g$; the bulk will be given by the appropriate asymptotically $AdS_3$ manifold $\mathcal{M}_3$, while the whole four-manifold is given by $\mathcal{M}_3\times S^1$. Parametrizing the circle with the angular coordinate $\theta$, we have a topological interface at the point $\theta=0 \sim \theta=2\pi$,
\begin{equation}\label{TopInt}
    S_{\text{int}}= \frac{N}{2\pi}\int A_UdA_L,
\end{equation}
which relates the boundary values of the connection on the two sides of the cut. Note that there are no four-dimensional terms in the action, so the defect term compasses the whole action. We are therefore considering the partition function
\begin{eqnarray}
    \mathcal{Z}_{\text{flat}} = \int_{F=0}\mathcal{D}A\,e^{iS_{\text{int}}[A_U, A_L]},
\end{eqnarray}
where $A$ is the flat four-dimensional connection, and $A_U$, $A_L$ are the boundary values. Notice, however, that due to the flatness condition $A_U=A_D$, and more generally, for any $\gamma\times\{\theta\}\subset M_3 \times S^1$, the holonomy along $\gamma$ does not depend on $\theta$. Therefore, we can restrict the path integration to be over the three-dimensional flat connections: 
\begin{eqnarray}
    \mathcal{Z}_{\text{flat}} = \int_{F=0}\mathcal{D}A_3\,e^{iS_{\text{int}}[A_3,\,A_3]}.
\end{eqnarray}
We conclude that the low-energy dynamics on $\mathcal{M}_3$ is governed by an appropriate Chern-Simons theory; this conclusion was first drawn in \cite{Ganor:2008hd} (see also \cite{Ganor:2010md, Ganor:2012mu}) for the case of self-dual coupling, where the IR TQFT is $U(1)_2$.

This conclusion can also be retrieved from the equations of motion discussed above. Indeed, the equation of motion \eqref{Eq:3DEOMs} in the large distance limits reduces to
\begin{equation}
    \partial_{\theta}\,F = \partial_{\theta}\,G=0,
\end{equation}
and together with the boundary conditions \eqref{Eq:DualityB.c.} leads to the solution\footnote{In fact, $F=G=0$ solves the system even before taking the large distance limit.}
\begin{equation}
    F = G = 0.
\end{equation}
In the presence of a Wilson line insertion, we assume the line to be localized on the cut (and recall that this assumption can be done without loss of generality); then the e.o.m. has the same large distance limit as above. Considering the $\mathcal{D}^{(2)}_N$ defect as an example, a Wilson line gives rise to the modified boundary conditions \eqref{Eq:DualityWilsonLowerB.c.}, \eqref{Eq:DualityWilsonUpperB.c.}; the modified solution is found to be 
\begin{equation}
    F=-\frac{2\pi n_L}{2N}\delta(\gamma_L),\qquad G=\frac{2\pi n_L}{2N}\delta(\gamma_L)
\end{equation}
in the \eqref{Eq:DualityWilsonLowerB.c.} case, and
\begin{equation}
    F=-\frac{2\pi n_U}{2N}\delta(\gamma_U),\qquad G=-\frac{2\pi n_U}{2N}\delta(\gamma_U)
\end{equation}
in the \eqref{Eq:DualityWilsonUpperB.c.} case. Note that for $n_U=n_L$ and $\gamma_U=\gamma_L$ these two solutions are related by the $\mathcal{S}$ transformation, and so physically equivalent, as expected.

\section{Lines and the near-the-boundary limit}

In the previous section, we have argued that the effective theory in AdS$_3$ consists of a tower of massive vector fields, as well as from a topological sector given by a Chern-Simons theory: this CS theory is obtained from the defect Lagrangians, by identifying the gauge fields on both sides of the cut. It follows that the large distance dynamics of the line operators is topological, and, in particular, this is the case for the near-the-boundary limit. In the original $\mathbb{R}^4$ formulation this corresponds to sending a line close to the monodromy defect. Let us consider in more detail the specific cases at hands.

\paragraph{$\mathcal{D}^{(2)}_N$ defect.} In this case the IR TFT is given by $U(1)_{2N}$ CS theory. At any $N\in\mathbb{N}$, it admits the chiral boundary condition, where the boundary carries a left-moving sector of the compact boson CFT. The lines can end at the boundary, and the end points correspond to insertions of chiral primary operators. Lines can also be sent to the boundary, giving rise to topological defects known as Verlinde lines. From the $\mathbb{R}^4$ perspective, it means that the monodromy defect carries a chiral sector. Moreover, all the bulk lines can end on the monodromy defect.

The Wilson-'t Hooft lines flow to the CS lines at large distance. It follows from \eqref{Eq:WilsonBulkDefect} that a Wilson line of charge $n$ is mapped to a CS line of the same charge (mod $2N$), giving rise to $2N$ distinct lines. In particular, the lines with charge a multiple of $2N$ become trivial in the IR, and can be completely absorbed when they are pushed to the defect.  

The mapping for the 't Hooft lines follows from \eqref{Eq:tHooftBulkDefect}, and implies that the lines with an odd magnetic charge are mapped to the line of charge $N$, while the lines with even magnetic charge become trivial in the IR.

As we have already mentioned, sending a line to the monodromy defect, we obtain a topological line on the defect. Therefore, we can consider the enriched defect $\mathcal{M}^{(2),\,\alpha}_{N}$, where the cohomology class $\alpha\in H^1(\Sigma,\,\mathbb{Z}_{2N})$ encodes the position of topological lines on the defect, and $\Sigma$ is the manifold, where the defect is supported. The procedure of sending a bulk line to the defect can be described by the fusion-like rules:
\begin{eqnarray}
    \mathcal{W}_n(\gamma)\times\mathcal{M}^{(2),\,\alpha}_{N} &=& \mathcal{M}^{(2),\,\alpha+n[\delta(\gamma)]}_{N},\\
    \mathcal{T}_m(\gamma)\times\mathcal{M}^{(2),\,\alpha}_{N} &=& \mathcal{M}^{(2),\,\alpha+m\,N[\delta(\gamma)]}_{N}.
\end{eqnarray}

\paragraph{$\mathcal{D}^{(2)}_{N_e, N_m}$ defect.} The IR TQFT is obtained from \eqref{Eq:Duality_Fractional} by identifying $A$ and $\tilde{A}$, and is given by
\begin{equation}
    \frac{1}{2\pi}\int N_m\,adb-N_e\,adA+bdA.
\end{equation}
The defect chiral sector is described by the set of chiral scalars with the same $K$-matrix, and includes $|\det K|=2N_mN_e$ primary operators.

Taking advantage of the fact that $N_e$ and $N_m$ are coprime, we can find two integers $p,\,q\in\mathbb{Z}$, such that $p\,N_m-q\,N_e=1$. The $SL(3, \mathbb{Z})$ transformation
\begin{equation}
    G=\left(
    \begin{matrix}
        -p\,q && p && q\\
        1 && 0 && 0\\
        -p\,N_m -q\,N_e && Ne && Nm
    \end{matrix}
    \right)
\end{equation}
brings the $K$-matrix to the form
\begin{eqnarray}
    K'\equiv G\cdot K\cdot G^T=
    \left(
    \begin{matrix}
        0 && 1 && 0\\
        1 && 0 && 0\\
        0 && 0 && 2N_e N_m
    \end{matrix}
    \right),
\end{eqnarray}
which reveals the sector corresponding to a trivial TQFT (a level $1$ BF theory), and the $U(1)_{2N_m N_e}$ CS theory. The new basis of the gauge fields is given by
\begin{eqnarray}
    c_1&=&-N_e\,A+N_m\,b,\\
    c_2&=&a+p^2\,N_m\,A+q^2\,N_e\,b,\\
    c_3&=&p\,A-q\,b,
\end{eqnarray}
where $c_1$ and $c_2$ correspond to the trivial factor, while $c_3$ corresponds to the non-trivial $U(1)_{2N_eN_m}$ sector.

In this new basis the IR images of the Wilson-'tHooft lines are found to be
\begin{eqnarray}
    \mathcal{W}_n&\sim& e^{i\int\,n_1\,c_1+n_3\,c_3},
    \quad n_1=q\,n_A+p\,n_b,\quad n_2 = n\,N_m,\nonumber\\
    \mathcal{T}_m&\sim&e^{i\int\,m_1\,c_1+m_2\,c_2+m_3\,c_3},\quad m_1=q\,n_A-n_a\,p\,q\,(p\,N_m+q\,N_e),\quad m_2=n_a,\\
    &&m_3=n\,n_A\,N_m-n_a(p^2\,N_m+q^2\,N_e)=m\,N_e\,\,\,\text{mod}\,2N_eN_m.
\end{eqnarray}

In fact, we can mode out the $c_1$ and $c_2$ fields, and taking $w=\exp(i\int c_3)$, the Wilson lines are identified as $\mathcal{W}_n\sim w^{nN_m}$ and the 't Hooft lines are identified as $\mathcal{T}_m\sim w^{mN_e}$. It follows that only Wilson lines of charge $0, \pm1, ...,N_e$ source a non-trivial primary operator while ending on the monodrom defect, and give rise to a non-trivial topological defect while brought on the defect. An equivalent conclusion is made for the 't Hooft lines, with $N_e$ replaced by $N_m$. Considering again the monodromy defects enriched with topological lines $\mathcal{M}^{(2),\,\alpha}_{N_e,\,N_m}$, where now the cohomology class $\alpha\in H^1(\Sigma,\,\mathbb{Z}_{2N_eN_m})$, we can write down the relations
\begin{eqnarray}
    \mathcal{W}_n(\gamma)\times\mathcal{M}^{(2),\,\alpha}_{N_e,\,N_m} &=& \mathcal{M}^{(2),\,\alpha+nN_m[\delta(\gamma)]}_{N_e,\,N_m},\\
    \mathcal{T}_n(\gamma)\times\mathcal{M}^{(2),\,\alpha}_{N_e,\,N_m} &=& \mathcal{M}^{(2),\,\alpha+nN_e[\delta(\gamma)]}_{N_e,\,N_m}.
\end{eqnarray}

\paragraph{$\mathcal{D}^{(3)}_{N}$ defects.} Here the effective IR TQFT is given by the $U(1)_{N}$ CS theory; for simplicity, we will consider the case of even level, which does not require a spin structure. As in the previous examples, the Wilson lines $\mathcal{W_n}$ are mapped to the CS lines of the same charge. Correspondingly, their ends induce the charges $0, \pm1,..., N/2$, and when brought to the monodromy defects, they generate $N$ independent topological lines. On the pther hand, it follows from the second relation of \eqref{Eq:TrialitytHooftFusion} that 't Hooft lines are mapped to a trivial line. This can be summarised in the following relations:
\begin{eqnarray}
    \mathcal{W}_n(\gamma)\times\mathcal{M}^{(3),\,\alpha}_{N} &=& \mathcal{M}^{(3),\,\alpha+n[\delta(\gamma)]}_{N},\\
    \mathcal{T}_m(\gamma)\times\mathcal{M}^{(3),\,\alpha}_{N} &=& \mathcal{M}^{(3),\,\alpha}_{N},
\end{eqnarray}
with $\alpha\in H^{(1)}(\Sigma,\,\mathbb{Z}_{N})$.

\section{Conclusions and Outlook}

In this note we explore monodromy defects for non-invertible symmetries in $4d$ Maxwell theory, using the mapping to $AdS_3\times S^1$. This allowed us to extract the spectrum of the defect primary operators.

One of our objectives was to study the behaviour of Wilson/'t Hooft operators in the presence of the monodromy defect. Due to the fact that the effective theory in $AdS_3$ is gapped, the lines become topological when sent close to the boundary (correspondingly, close to the defect), and their behaviour is governed by a Chern-Simons theory. Similarly to what happens in a CS theory with a boundary, the lines can end on the boundary (on the defect), and the end points correspond to chiral primaries of the boundary (defect) theory.

Some interesting questions remain open. First, the holographic approach allows for the computation of some other DCFT observables, such as the defect central charge \cite{Giombi:2021uae}. Second, we have discussed only the simplest instances of non-invertible topological defects: the duality defects at $\tau=\frac{iN_e}{N_m}$ and the triality defects at $\tau=e^{i\pi/3}N$. On the other hand, it was shown in \cite{Niro:2022ctq} that similar topological defects can be found at any rational value of $\tau$: it would be of interest to generalize our analysis to these more general cases.

Finally, one may proceed with other models enjoying non-invertible symmetries, with $\mathcal{N}=4$ SYM being a prominent example. In fact, $\mathcal{S}$-duality transformations act on supercharges \cite{Kapustin:2006pk}, so such a monodromy defect would break supersymmetry completely. In order to preserve some of it, one also has to turn on monodromy for the $R$-symmetry: in this way one can preserve up to $12$ supercharges, giving rise to $\frac{3}{4}$ BPS monodromy defects \cite{Ganor:2008hd, Kaidi:2022uux}. Using again the mapping the mapping to $AdS_3\times S^1$ with an twist. The question of $\frac{3}{4}$ BPS $\mathcal{S}$-duality-twisted compactifications for $U(N)$ $\mathcal{N}=4$ SYM was analysed in \cite{Ganor:2008hd} and, interestingly enough, it was found that the resulting $3d$ theory is gapped only for sufficiently small values of $N$. More precisely, this is the case for $n=1,2, 3$ at the duality point $\tau=i$ and $n=1,...,5$ for the triality point $\tau=e^{\frac{2\pi i}{3}}$. Correspondingly, the topological behaviour in the near-the-boundary (or near-the-defect) limit is expected only in these low-lying values of the rank of the gauge group.

\bibliographystyle{JHEP}
\bibliography{biblio}

\end{document}